\documentclass[
]{ceurart}

\usepackage{listings}
\begin{document}

\copyrightyear{2026}
\copyrightclause{Copyright for this paper by its authors.
  Use permitted under Creative Commons License Attribution 4.0
  International (CC BY 4.0).}

\conference{In Proceedings of the ACM CHI 2026 Workshop on Shaping Future Human Connections: Social Augmentation through XR Technologies, April 13, 2026, Barcelona, Spain.}

\title{Visual-to-Haptic Augmentation in XR: A Wearable Glove for Perceptual Grounding in Multimodal Interaction}
\author{Mohd Faisal}[%
orcid=0000-0003-3363-4758,
email=mmohd055@uottawa.ca
]
\cormark[1]

\author{Hamdi Elsaddik}[%
email=helsaddi@uottawa.ca
]

\author{Erhan Baturay Onural}[%
email=eonur052@uottawa.ca
]

\author{Jihong Zhang}[%
email=jzhan554@uottawa.ca
]

\author{Fedwa Laamarti}[%
email=flaamart@uottawa.ca
]

\author{A.El Saddik}[%
email=elsaddik@uottawa.ca
]

\address{School of Electrical Engineering and Computer Science, University of Ottawa, Ottawa, Ontario, Canada}

\cortext[1]{Corresponding author.}

\begin{abstract}
Extended Reality (XR) systems increasingly deliver high-fidelity visual and auditory experiences, yet tactile perception remains comparatively underutilized as a modality for enriching embodied interaction. This work presents a visual-to-haptic wearable glove and a feature-based visual-to-haptic mapping algorithm that translates spatial and temporal visual features from images and videos into distributed vibrotactile patterns. The proposed method extracts motion, edge, and brightness cues and fuses them into actuator-level intensity maps aligned with a 29-actuator glove arranged in a five-by-seven layout.

The system is implemented through a modular four-layer architecture comprising the XR environment, media content handling, visual-to-haptic processing, and embedded haptic hardware. A within-subject user study (N = 20) compared visual-only interaction with visual-plus-haptic augmentation across texture-based and dynamic video scenarios. Results indicate that tactile augmentation significantly improves perceived realism in dynamic video scenarios and enhances immersion and visual–tactile correspondence across conditions, with stronger and more consistent effects observed for dynamic visual events.

While the current implementation operates in a single-user, offline-synchronized configuration, the findings demonstrate that vision-driven tactile augmentation can function as a perceptual enhancement layer within multimodal XR systems. Such a layer may provide a foundation for future socially enriched XR environments where coherent multisensory grounding supports higher-level interaction and communication.
\end{abstract}


\maketitle

\section{Introduction}
Extended Reality (XR) systems increasingly support immersive interaction across entertainment, education, training, and collaborative scenarios. While contemporary XR platforms deliver high-fidelity visual and auditory experiences, tactile feedback remains comparatively underrepresented, despite its fundamental role in human perception and embodiment \cite{r1n, r2n}. This imbalance creates a perceptual gap between what users see and what they physically sense, limiting multimodal coherence in immersive environments.

Wearable haptic interfaces, particularly glove-based systems, have been investigated as a means of restoring tactile feedback in XR \cite{r4n}. Our prior work also explored a vibrotactile glove for texture and dynamic event simulation in VR \cite{r3n}. Advances in lightweight actuators and textile integration have made vibrotactile gloves practical for immersive applications \cite{r5n, r6n}, and tactile cues have been shown to enhance immersion and realism during virtual interaction \cite{r1n, r7n}. However, many existing approaches rely on pre-authored haptic libraries or predefined mappings, restricting adaptability to diverse and dynamic visual content.

Research on multimodal interaction demonstrates that combining visual and tactile cues strengthens embodied perception \cite{r35}. In parallel, work on social augmentation in XR highlights the importance of enriching perceptual cues to support engagement and inclusivity \cite{r10n, r11n}. While these efforts largely emphasize visual and auditory channels, they ultimately depend on coherent perceptual grounding across modalities. Tactile augmentation therefore represents a complementary pathway for reinforcing multisensory coherence in XR.

Visual-to-haptic rendering provides a mechanism for translating visual information directly into tactile stimulation. Prior studies have mapped image intensity and structural visual features to vibrotactile patterns to simulate surface textures and cross-modal tactile effects \cite{r12n, r13n}, yet many systems address static and dynamic content separately, or rely on manually authored effects.

In this work, we propose a feature-based visual-to-haptic mapping algorithm implemented within a wearable glove system that translates spatial and temporal visual features into distributed tactile feedback. The hardware platform extends our earlier haptic glove prototype for immersive VR texture and dynamic event simulation \cite{r3n}, and introduces a unified visual-to-haptic mapping algorithm together with a structured perceptual evaluation framework. Images and videos are converted into actuator-level intensity maps delivered across a 29-actuator glove arranged in a 5 $\times$ 7 layout. The system follows a four-layer architecture comprising the XR environment, media handling, visual-to-haptic processing, and embedded hardware layers.

A controlled within-subject study compared visual-only interaction to visual-plus-haptic augmentation across texture-based and dynamic scenarios. Results indicate that tactile augmentation improves perceived immersion, realism, and visual--tactile correspondence, with stronger effects observed for dynamic events. Although the current implementation operates in a single-user, offline-synchronized configuration, the findings suggest that visual-to-haptic augmentation can serve as a perceptual enhancement layer within multimodal XR systems and may support socially enriched experiences grounded in coherent multisensory interaction.

In this work, perceptual augmentation refers to the systematic enhancement of sensory coherence across modalities, where visual structure is reinforced through aligned tactile feedback. Rather than introducing new semantic or social signals, the proposed system strengthens the perceptual grounding of existing visual content, thereby supporting higher-level interaction in XR.

The remainder of this paper is structured as follows. Section~2 reviews related work. Section~3 presents the system and mapping algorithm. Section~4 describes the experimental evaluation. Section~5 discusses implications and limitations.

\section{Related Work}

Research on immersive XR increasingly emphasizes multisensory interaction, particularly the integration of tactile feedback alongside visual and auditory channels. Prior work spans wearable haptics, visual-to-haptic rendering, multimodal perception, and socially mediated touch.

\subsection{Haptic Wearables and Tactile Augmentation}

The sense of touch plays a fundamental role in perception and embodiment, yet remains underrepresented in mainstream XR systems \cite{r1n, r2n}. Wearable haptic devices, especially gloves, have been explored to restore tactile feedback \cite{r3n, r4n}. Advances in lightweight actuators and textile integration have enabled increasingly practical wearable haptic gloves for immersive XR interaction \cite{r5n, r6n}. Pneumatic and electrostatic approaches further expand tactile expressivity \cite{r4n, r6n}, while textile-based systems embed vibrotactile elements directly into wearable fabrics to enhance comfort and spatial resolution \cite{r36}.

Empirical studies report that tactile cues improve perceived immersion and realism \cite{r4n}, and psychophysical studies have shown that vibration amplitude and frequency influence perceived surface roughness and texture granularity \cite{r15}. Visual–haptic congruence has been shown to enhance engagement and perceptual coherence in multisensory experiences \cite{r35}, yet many wearable systems still depend on manually authored vibration libraries, limiting adaptability to dynamic visual content.

\subsection{Visual-to-Haptic Rendering}

Visual-to-haptic rendering translates image and video content into tactile stimulation. Early approaches mapped pixel intensity and structural image features directly to actuator-level vibration patterns \cite{r27}. Later work incorporated visual saliency and temporal feature extraction to improve realism in image- and video-driven haptic systems \cite{r28}. Dynamic visual events such as motion and impacts have been encoded into time-varying vibrotactile feedback through multimedia effect generation frameworks \cite{r28, r32, r33}.

More recently, machine learning methods have been proposed to learn visual-to-tactile mappings using cross-modal generation models \cite{r29, r30, r38}. Although these approaches enable complex nonlinear mappings, they typically require large paired datasets and introduce computational overhead that can limit real-time wearable deployment.

Despite these advances, existing systems often treat static texture rendering \cite{r15} and dynamic event translation \cite{r28, r33} separately, or rely on manually authored mappings \cite{r32}. There remains a need for a unified wearable approach that integrates spatial and temporal visual features within a single, interpretable real-time pipeline.

\subsection{Multimodal Perception and Social XR Context}
Human perception integrates visual and tactile inputs into coherent multisensory experiences. Cross-modal studies demonstrate that synchronized tactile cues enhance engagement and perceived realism \cite{r35}. In distributed XR systems, however, latency, jitter, and transmission instability may disrupt perceptual coherence and degrade immersion \cite{r5, r48}. Reviews of haptic network protocols further emphasize the importance of ultra-low-latency transmission and adaptive synchronization strategies to preserve multisensory alignment in immersive environments \cite{r48}. Pseudo-haptic effects further illustrate how visual cues influence tactile perception, reinforcing the importance of consistent cross-modal grounding \cite{r35}.

Beyond perceptual integration, recent work in social XR highlights the role of enriched sensory cues in supporting engagement, trust, and mediated interaction \cite{r10n, r11n}. Broader analyses of metaverse-scale XR systems identify multimodal coherence, system integration complexity, and perceptual alignment as central technical challenges for immersive environments \cite{r17, r51}. Maintaining consistent cross-modal grounding across visual, auditory, and emerging sensory channels is therefore critical for scalable and socially engaging XR experiences.

Systems for mediated touch and expressive haptics demonstrate how vibrotactile signals can convey affect and social intent \cite{r32, r33, r41, r42, r43}. Recent work on digital twin robotic systems further shows that physically grounded haptic interaction enhances social presence in metaverse environments. Remote handshake systems implemented with anthropomorphic robotic arms demonstrate that bidirectional haptic feedback strengthens embodied co-presence and interaction realism \cite{r47}. These efforts depend fundamentally on reliable multisensory synchronization and adaptable haptic interfaces.

Despite substantial progress across hardware, rendering algorithms, and social XR research, existing solutions remain fragmented. Many systems focus on either static texture rendering \cite{r15} or dynamic event translation \cite{r28, r33}, or rely on manually authored mappings \cite{r32}, limiting adaptability. There remains a need for a practical wearable system that integrates visual-to-haptic translation for both spatial and temporal content within a unified XR pipeline. The present work addresses this gap by proposing a feature-based mapping algorithm within a modular glove-based architecture.

\section{System Design and Implementation}

\begin{figure}[htbp]
\centering
\includegraphics[width=\linewidth]{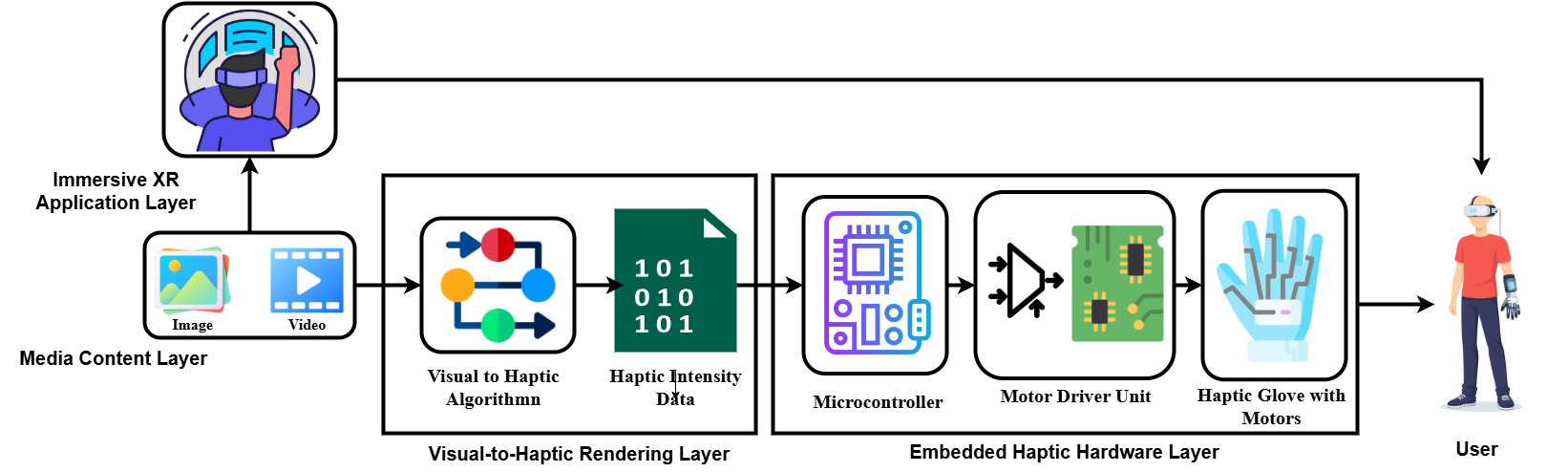}
\caption{Four-layer architecture of the proposed visual-to-haptic augmentation framework, illustrating the flow from XR visual content to actuator-level tactile output.}
\label{fig:architecture}
\end{figure}

The proposed system delivers visual-to-haptic augmentation through a modular four-layer architecture: (i) Immersive XR Application Layer, (ii) Media Content Layer, (iii) Visual-to-Haptic Rendering Layer, and (iv) Embedded Haptic Hardware Layer (Figure~\ref{fig:architecture}). These layers collectively translate spatial and temporal visual features into distributed vibrotactile feedback delivered via a wearable glove comprising 29 actuators arranged in a $5 \times 7$ spatial layout.

The system operates in a frame-based processing mode at 150\,ms intervals. This sampling rate was selected to balance perceptual responsiveness with signal stability and actuator limitations. Although currently evaluated in a single-user configuration, the modular structure enables integration into broader XR pipelines, including socially enriched and multi-user environments requiring multisensory grounding.

\subsection{Immersive XR Application Layer}

\begin{figure}[htbp]
\centering
\includegraphics[width=\linewidth]{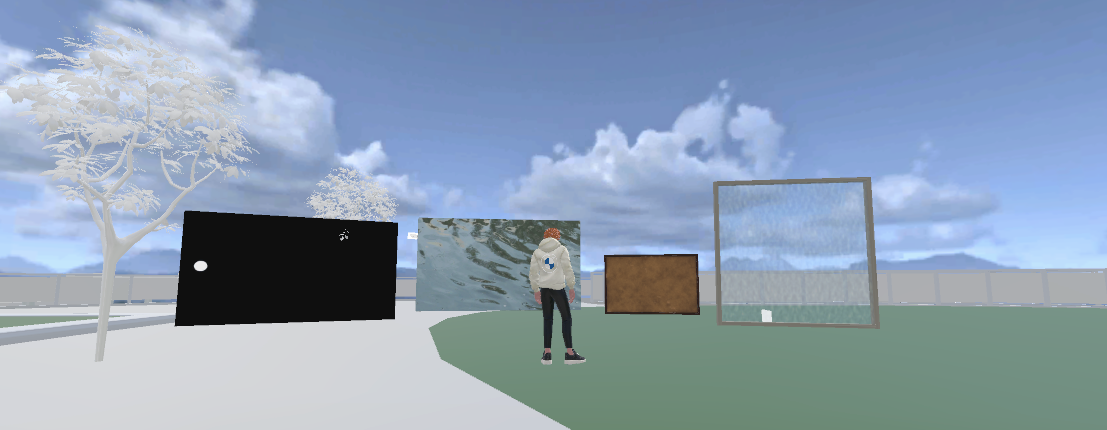}
\caption{Example XR scene containing textured objects and embedded video panels used to generate visual-to-haptic feedback.}
\label{fig:vr_scene}
\end{figure}

This layer generates the visual stimuli that serve as input for tactile translation. The XR environment is implemented in Unity and presents textured virtual objects and embedded video panels designed to evoke tactile interpretations through surface properties and dynamic motion cues (Figure~\ref{fig:vr_scene}).

Visual frames from the rendered scene are streamed to the Media Content Layer for analysis. The current prototype operates in a desktop-based XR configuration; however, the architecture is compatible with VR head-mounted displays and collaborative XR contexts.

\subsection{Media Content Layer}

The Media Content Layer manages visual assets embedded within the XR scene, including static images and dynamic video sequences. It forwards raw frames to the Visual-to-Haptic Rendering Layer without modifying visual content.

Two processing pathways are activated depending on input type:

\begin{itemize}
\item \textbf{Static Image Mode}: A single frame is processed to generate one spatial intensity map.
\item \textbf{Dynamic Video Mode}: Consecutive frames are sampled at 150\,ms intervals to generate temporally indexed tactile sequences.
\end{itemize}

This separation ensures consistent handling of both spatial textures and time-varying events within a unified rendering framework.

\subsection{Visual-to-Haptic Rendering Layer}

The Visual-to-Haptic Rendering Layer constitutes the core algorithmic contribution of this work. We propose a feature-based visual-to-haptic mapping algorithm that translates visual structure into actuator-level vibration intensities aligned with the glove geometry.

\subsubsection{Grid Mapping and Preprocessing}

\begin{figure}[htbp]
\centering
\includegraphics[width=0.75\linewidth]{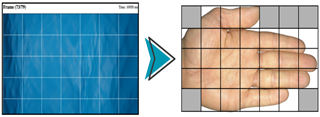}
\caption{Spatial correspondence between the $5 \times 7$ visual grid and the 29-actuator glove layout. Gray cells indicate non-mapped regions.}
\label{fig:grid_mapping}
\end{figure}

Each visual frame is segmented into a $5 \times 7$ grid corresponding spatially to the 29 actuators on the glove (Figure~\ref{fig:grid_mapping}). Six peripheral grid cells are excluded to accommodate ergonomic constraints.

Before feature extraction, frames undergo:

\begin{itemize}
\item Grayscale conversion,
\item Gaussian smoothing to reduce high-frequency noise,
\item Optional downscaling to ensure computational stability.
\end{itemize}

These preprocessing steps improve robustness against minor visual fluctuations that could otherwise trigger unintended tactile activation.

\subsubsection{Feature Extraction}

For each grid cell, three perceptually motivated features are computed:

\begin{itemize}
\item \textbf{Optical Flow Magnitude ($F$)}: Estimated using the dense Farnebäck method to capture motion intensity between consecutive frames.
\item \textbf{Sobel Edge Magnitude ($S$)}: Detects structural transitions and surface boundaries.
\item \textbf{Inverted Brightness ($B$)}: Highlights visually salient darker regions potentially corresponding to texture depth or contrast.
\end{itemize}

Each feature is normalized to the range $[0,1]$ to enable consistent fusion across modalities.

\subsubsection{Feature Fusion and Parameter Selection}

The normalized features are combined using a weighted linear fusion model:

\[
V = \alpha F + \beta S + \gamma B
\]

where $\alpha$, $\beta$, and $\gamma$ control the relative contribution of motion, edge structure, and brightness.

Weights were empirically tuned through pilot experiments to balance perceptual saliency and tactile clarity:

\begin{itemize}
\item For dynamic video: $(\alpha,\beta,\gamma) = (0.2, 0.4, 0.4)$,
\item For static images: $(0, 0.5, 0.5)$.
\end{itemize}

These weights were selected to balance perceptual saliency with actuator stability and hardware constraints, ensuring clear tactile differentiation without excessive vibration saturation. These values emphasize structural and luminance cues while maintaining responsiveness to motion in dynamic scenarios.

\subsubsection{Postprocessing and Temporal Stabilization}

To reduce spurious activations, intensity values below a threshold of 15 (on a 0–255 scale) are suppressed. This threshold was determined through preliminary calibration to remove low-amplitude noise while preserving perceptually meaningful tactile cues.

For video sequences, a three-frame moving average is applied to smooth temporal transitions and reduce abrupt actuator fluctuations.

Each processed frame produces a 29-element intensity array corresponding to actuator commands. Static images yield a single tactile map; video sequences generate time-indexed arrays aligned with frame sampling intervals.

\subsubsection{Visualization Dashboard}

\begin{figure}[htbp]
\centering
\includegraphics[width=5in]{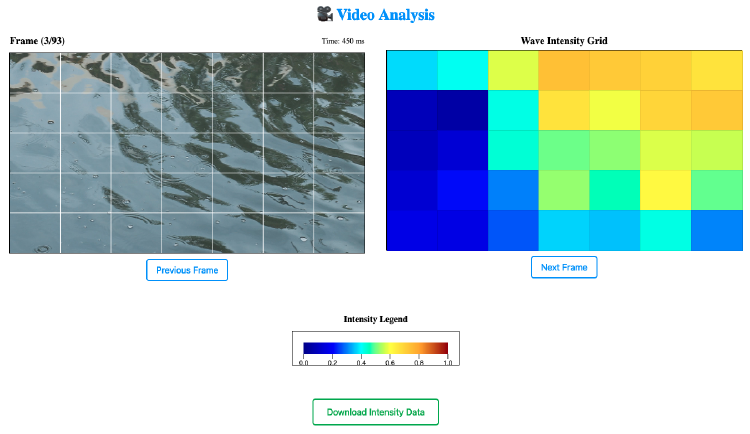}
\caption{Visualization dashboard showing a selected video frame and corresponding actuator intensity map.}
\label{fig:dashboard}
\end{figure}

To support inspection and reproducibility, a lightweight browser-based dashboard was developed to visualize actuator intensity distributions during development and testing (Figure~\ref{fig:dashboard}). The interface displays the visual frame alongside the corresponding $5 \times 7$ intensity grid using a color gradient to represent vibration magnitude. Actuator intensities are visualized using a perceptually ordered color gradient ranging from dark blue (low intensity) to red (high intensity), corresponding to normalized vibration magnitudes mapped from 0 to 255. This visualization aids qualitative inspection of spatial distribution and temporal evolution during calibration but does not alter the underlying rendering algorithm.

This tool facilitates qualitative validation of spatial mapping and temporal consistency during system calibration.

\subsection{Embedded Haptic Hardware Layer}

The haptic glove is a custom-developed prototype designed by the authors. It integrates 29 Linear Resonant Actuators (LRAs) embedded within stretchable textile to ensure comfort and consistent skin contact.

Each actuator is driven by a DRV2605L haptic driver connected through four I\textsuperscript{2}C multiplexers (addresses 0x70–0x73) to a Teensy 4.1 microcontroller. Firmware written in C++ parses actuator intensity arrays, maps values (0–255) to PWM duty cycles, and updates actuators at 150\,ms intervals to maintain alignment with visual frame processing.

Power is supplied via a regulated 5\,V, 2\,A distribution bus. A compact wrist-mounted enclosure houses the controller and Li-Po battery, with flexible silicone wiring supporting user mobility. The hardware architecture allows scalability through additional multiplexers or higher-density actuator configurations in future iterations.

\section{Experimental Setup and Evaluation}

\subsection{Participants and Apparatus}

Twenty participants (10 male, 10 female; aged 20–35) voluntarily participated in the study. All reported normal or corrected-to-normal vision and no prior experience with the proposed haptic system. The experimental setup consisted of a \textit{Meta Quest 2} VR headset paired with the custom-developed 29-actuator glove described in Section~3, driven by a \textit{Teensy 4.1} microcontroller and \textit{DRV2605L} haptic drivers.

Experiments were conducted in a quiet indoor environment to minimize external distractions. The study was approved by the University of Ottawa Research Ethics Board (H-09-23-9473). All participants provided written informed consent and responses were anonymized.

\subsection{Experimental Design}

A within-subject design was adopted to compare visual-only interaction with visual-to-haptic augmentation across two types of XR content. The first module involved interaction with virtual textured surfaces (glass representing smooth texture and sandpaper representing rough texture). The second module involved dynamic video content (flowing water and a bouncing ping-pong ball) designed to emphasize temporal visual features.

Each participant experienced both the visual-only condition and the visual + haptic condition for both modules. The order of conditions was counterbalanced to reduce learning and ordering effects. Each condition lasted approximately two minutes, after which participants completed a structured questionnaire. During haptic trials, tactile signals were synchronized with visual content using the system described in Section 3. Although the sample size (N = 20) is typical for within-subject XR studies, future work will include larger cohorts to confirm effect stability.

\subsection{Measures and Statistical Analysis}

Participants rated their experiences using five-point Likert scales (1 = strongly disagree, 5 = strongly agree). For both modules, perceived realism was collected under visual-only and visual + haptic conditions, enabling direct baseline comparison. Under the haptic condition, additional perceptual dimensions were assessed, including immersion, responsiveness/timing, tactile–visual correspondence, and engagement.

To evaluate the effect of haptic augmentation on realism, paired-sample t-tests were conducted comparing visual-only and visual + haptic conditions. Statistical significance was assessed at $\alpha = 0.05$. Effect sizes were computed using Cohen’s $d_z$ for repeated measures to quantify the magnitude of observed differences. Descriptive statistics (mean and standard deviation) were computed for all haptic-condition metrics.

\subsection{Results}

\subsubsection{Baseline Comparison: Realism}

For the texture module, no significant difference was observed between visual-only ($M=3.45$, $SD=1.39$) and visual + haptic ($M=3.50$, $SD=1.32$) conditions, $t(19)=0.11$, $p=0.914$, $d_z=0.03$. The negligible effect size indicates that haptic augmentation did not significantly alter perceived realism for static surface textures.

For the video module, haptic feedback significantly increased perceived realism compared to baseline ($M_{visual}=3.25$, $SD=1.33$; $M_{haptic}=3.95$, $SD=1.05$), $t(19)=2.15$, $p=0.044$, $d_z=0.48$. The moderate effect size suggests that dynamic visual events benefit substantially from tactile augmentation. The 95\% confidence interval for the mean difference did not include zero, further supporting the reliability of the observed effect.

\subsubsection{Perceptual Ratings Under Haptic Augmentation}

Table~\ref{tab:haptic_metrics} presents descriptive statistics for perceptual ratings under the visual + haptic condition. Across all perceptual dimensions, the video module yielded higher mean ratings and lower variability, indicating stronger temporal coherence and perceptual consistency when dynamic visual features were translated into tactile signals.

\begin{table}[h]
\centering
\caption{Mean $\pm$ SD ratings under the visual + haptic condition ($N = 20$).}
\label{tab:haptic_metrics}
\begin{tabular}{lcc}
\toprule
\textbf{Metric} & \textbf{Texture} & \textbf{Video} \\
\midrule
Realism & 3.50 $\pm$ 1.32 & 3.95 $\pm$ 1.05 \\
Immersion & 3.65 $\pm$ 1.23 & 4.10 $\pm$ 0.97 \\
Timing / Responsiveness & 3.90 $\pm$ 1.02 & 4.30 $\pm$ 0.86 \\
Tactile--Visual Correspondence & 3.15 $\pm$ 1.42 & 3.75 $\pm$ 0.97 \\
Engagement & 3.85 $\pm$ 1.14 & 4.30 $\pm$ 0.98 \\
\bottomrule
\end{tabular}
\end{table}

\subsection{Qualitative Observations}

Open-ended responses reinforced the quantitative findings. Participants frequently described video-based tactile feedback as “natural,” “timely,” and “immersive,” emphasizing strong temporal alignment between visual motion and vibration patterns. In contrast, texture feedback was generally described as distinguishable but occasionally discrete, suggesting that higher actuator density or refined spatial interpolation could further enhance surface continuity.

Overall, the results indicate that the proposed visual-to-haptic rendering approach is particularly effective for dynamic visual content, while static texture rendering may benefit from improved spatial granularity.

\section{Discussion}

The results indicate that the proposed visual-to-haptic mapping framework effectively translates dynamic visual features into perceptually meaningful tactile feedback in immersive XR. The statistically significant improvement in perceived realism for the video module confirms that motion- and structure-based feature fusion can support coherent cross-modal augmentation when temporal cues are present. The moderate effect size suggests a practically meaningful perceptual benefit for dynamic content.

For static textures, a numerical increase in realism was observed under the visual + haptic condition, though this difference did not reach statistical significance. This suggests that while the current actuator configuration supports perceptible differentiation between smooth and rough surfaces, spatial granularity may limit fine surface continuity. These constraints relate primarily to hardware resolution rather than the feature-based mapping strategy itself. Future iterations will explore higher-density actuator layouts and spatial interpolation strategies to improve continuous texture rendering while maintaining ergonomic feasibility.

Across perceptual metrics, the video module consistently yielded higher ratings and lower variability, indicating stronger temporal coherence when motion-based visual cues were translated into vibration patterns. This supports prior findings in multimodal perception that temporally aligned sensory signals enhance perceptual binding and immersion.

The feature-based mapping approach provides interpretability and low computational overhead compared to data-driven cross-modal synthesis methods, making it suitable for real-time embedded deployment within wearable XR systems.

\subsection{Implications and Future Directions}

The findings support interpreting visual-to-haptic translation as a perceptual augmentation layer in XR. Rather than introducing new semantic or socially encoded signals, the system reinforces existing visual structure through aligned tactile feedback, strengthening multisensory grounding.

Although evaluated in a single-user configuration, such perceptual grounding may contribute to improved social coordination, communication clarity, and collaborative task performance in multi-user XR environments. Future work will focus on automated event-driven synchronization, optimization of actuator density and spatial distribution, and extended studies incorporating task-based performance measures.

Overall, the proposed framework establishes a practical and interpretable foundation for tactile augmentation in XR, particularly for dynamic visual content.

\section{Conclusion}

This work presented a feature-based visual-to-haptic glove system capable of rendering both static textures and dynamic visual events in XR. A within-subject study ($N=20$) showed statistically significant improvements in perceived realism for dynamic video scenarios when haptic feedback was included, while texture differentiation was perceptible but not significantly improved relative to baseline.

These findings validate the proposed mapping framework for dynamic cross-modal augmentation and highlight opportunities for enhanced spatial rendering through increased actuator density. The system establishes a practical foundation for perceptual augmentation in XR, supporting future integration into interactive and socially enriched immersive environments. Such perceptual grounding may contribute to more coherent and inclusive socially augmented XR environments.

\section{Acknowledgements}
This work has been partially supported by the Natural Sciences and Engineering Research Council of Canada (NSERC). Erhan Baturay Onural is supported by the Republic of Türkiye through the YLSY Study Abroad Programme.
\bibliography{sample-ceur}

\appendix

\end{document}